\documentclass[twocolumn,pra,superscriptaddress,showpacs]{revtex4}

\usepackage{dcolumn}
\usepackage{amsmath}
\usepackage{bm}
\begin{document}

\title{Spin-dipole induced lifetime of the least-bound
${}^{5}\Sigma^{+}_{g}$ state of
He(2${}^{3}\text{S}_{1}$)+He(2${}^{3}\text{S}_{1}$)}
\author{Timothy J. Beams}
\affiliation{School of Mathematical and Physical Sciences,
James Cook University, Townsville, Australia 4811}
\author{Gillian Peach}
\affiliation{Department of Physics and Astronomy, University College London,
Gower Street, London, WC1E 6BT, United Kingdom}
\author{Ian Whittingham}
\affiliation{School of Mathematical and Physical Sciences,
James Cook University, Townsville, Australia 4811}

\date{\today}

\begin{abstract}
The properties of the least-bound vibrational level ($v=14$) of the
${}^{5}\Sigma^{+}_{g}$ state formed during the ultracold collision of
two spin-polarized metastable 2${}^{3}\text{S}_{1}$
helium atoms are crucial to studies of
photoassociation spectroscopy of metastable helium. We report a
calculation of the autoionization lifetime $\tau_{g}$
of this state induced by
spin-dipole coupling of the ${}^{5}\Sigma^{+}_{g}$  state to the
${}^{1}\Sigma^{+}_{g}$ state from which Penning and associative ionization
processes are highly probable. We find $\tau_{g} \approx 150\;\mu$s,
significantly larger than the recent experimental estimates of
$(4-5)\;\mu$s.

\end{abstract}

\pacs{32.70.Jz, 32.80.Pj, 32.80.Dz, 34.20.Cf}
\maketitle

\section{Introduction}

Two-photon photoassociation spectroscopy of ultracold spin-polarized
metastable 2${}^{3}\text{S}_{1}$ helium atoms
has recently been used \cite{Moal06} to
accurately measure the binding energy of the least bound vibrational level
($v=14$) of the ${}^{5}\Sigma^{+}_{g}$ state formed during the collision
of two atoms. The measured binding energy $E_{b}(v=14) =91.35 \pm 0.06$ MHz,
combined with the new  \textit{ab initio} ${}^{5}\Sigma^{+}_{g}$ potential of
Przybytek and Jeziorski \cite{Przyb05}, yielded the high precision value
$a=7.512 \pm 0.005$ nm for the $s$-wave scattering length.

Knowledge of the decay width $\gamma_{g}$ of this bound
state is important to the analysis of several experimental techniques used to
study photoassociation in metastable helium. In two-photon photoassociation
\cite{Moal06,Koel04}, for the case where one laser is detuned far from
resonance, the width of the narrow stimulated Raman line is determined
largely by $\gamma_{g}$, and, in dark resonance experiments, the incomplete
suppression of the photoassociation signal at the
resonance minimum is due to $\gamma_{g}$. From
these two experiments Moal \textit{et al} \cite{Moal06} obtain the estimate
$0.05 < \gamma_{g}/2\pi < 0.3$ MHz, corresponding to a lifetime $\tau_{g}$
between 0.5 and 3 $\mu $s. The decay width is also needed in the
interpretation of one-laser photoassociation experiments using calorimetric
detection \cite{Leonard03,Leonard04,Kim04} as decay of the $v=14$ bound
state contributes to the heating of the trapped atoms.

Koelemeij and Leduc \cite{Koel04} have given a theoretical estimate of
$\tau_{g}$ based upon the semiclassical vibration frequency of the $v=14$
level and inelastic decay rates due to spin relaxation and relaxation-induced
ionization \cite{Fedichev96,Venturi99}. They obtain $\tau_{g} =(4-5)\;\mu$s,
in good agreement with the measured linewidths.
However, these inelastic decay rates
were calculated for initial scattering states in the ${}^{5}\Sigma^{+}_{g}$
potential, not for a bound vibrational state. In this paper we shall
calculate the decay of the $v=14$ bound state due to relaxation-induced
ionization and argue that there is no contribution from spin-relaxation.

\section{Perturbation theory}

A collision between two atoms with spin quantum numbers $S_{1}$ and $S_{2}$
involves the total spin state $|(S_{1}S_{2})SM_{S}\rangle$ where $M_{S}$ is
the projection of $S$ onto the molecular axis. For metastable helium
atoms $S_{1}=S_{2}=1$ and Penning and associative ionization processes
are forbidden from the spin-polarized $|(11)22\rangle $ quintet state.
However, the spin-dipole interaction between the metastable helium atoms
couples the spin-polarized
$|SM_S\rangle = |22 \rangle $ state to the $|21 \rangle $, $|20 \rangle $ and
$|00 \rangle $ electronic spin states.  As Penning and
associative ionization from the $S=0$
singlet state are highly probable, the spin-dipole induced coupling
results in a finite lifetime of the $|22 \rangle $ state.  This
spin-dipole interaction can be treated as a perturbation and the method
presented here is a adaptation of that given in \cite{Beams05}.

The total Hamiltonian for the relative motion is
\begin{equation}
\label{eq1.1}
H = H_0 + H_{\text{sd}},
\end{equation}
where the Hamiltonian $H_{0}$ for the unperturbed system is
\begin{equation}
\label{eq1.2}
H_{0} = -\frac{\hbar^{2}}{2M}\;\nabla^{2}_{r}+ V^{\text{el}}_{\Lambda S}(r).
\end{equation}
Here $M$ is the reduced mass of the system and the adiabatic potential for
the molecular state ${}^{2S+1}\Lambda $ formed during the collision is
denoted by $V_{\Lambda S}(r)$ where $\Lambda $ is the projection of the
total electronic orbital angular momentum along the molecular axis.
The spin-dipole interaction is
\begin{equation}
\label{eq1.3}
H_\text{sd} = - \frac{\beta}{\hbar^2 r^3} \left[ 
3(\bm{S}_{1}\cdot \hat{\bm{r}})(\bm{S}_{2}\cdot \hat{\bm{r}})
          - \bm{S}_{1} \cdot \bm{S}_{2} \right],
\end{equation}
where $\bm{S}_1$ and $\bm{S}_2$ are the electronic-spin
operators for the two atoms, $\hat{\bm{r}} = \bm{r}/r$ is a unit vector
directed along the internuclear axis and
\begin{equation}
\label{eq1.4}
\beta = \alpha^2 \left( \frac{\mu_e}{\mu_B} \right)^2 E_h a_0^3,
\end{equation}
where $\alpha$ is the fine structure constant, $a_0$ is the Bohr radius,
$(\mu_e/\mu_B)$ is the electron magnetic moment to Bohr magneton ratio
and $E_h = \alpha^{2} m_e c^2 $ is the Hartree energy.

The eigenstates $|\alpha \rangle $ of $H_0$ can be  written
\begin{equation}
\label{eq1.5}
| \alpha \rangle = r^{-1}R_{va}(r)\;|\Phi_{a} \rangle
\end{equation}
where $\alpha =(v,a)$, and $a =\{\Gamma ,S, M_{S},l,m\}$  labels the
channel states
\begin{equation}
\label{eq1.6}
|\Phi_{a} \rangle = |\Gamma S M_{S} \rangle |lm \rangle 
\end{equation}
involving the relative motion states $|l m\rangle =Y_{lm}(\theta ,\phi)$.
The label $\Gamma $
represents the remaining quantum numbers $S_{1},S_{2}, \ldots $ needed to
fully specify the channel. The radial eigenfunction $R_{va}(r) $ satisfies
\begin{equation}
\label{eq1.7}
\left[ - \frac{\hbar}{2M} \frac{d^2}{dr^2} + \frac{l(l+1)\hbar^2 }{2M r^2} +
V^{\text{el}}_{\Lambda S}(r) \right] R_{va}(r) = E_\alpha^{(0)} R_{va}(r)
\end{equation}
where, for the initial state, $v$ labels the vibrational level.

The spin-dipole interaction may be written as the scalar product of two
second-rank irreducible tensors $H_{\text{sd}} = V_p(r) \bm{T}^2
\boldsymbol{\cdot} \bm{C}^2$, where $\bm{T}^2$ is
\begin{equation}
\label{eq1.8}
T^2_q \equiv \left[\bm{S}^1_1 \boldsymbol{\times}
\bm{S}^1_2 \right]^2_q = \sum_\mu C(1,1,2;\mu, q-\mu, q)\;
                        S^1_{1,\mu} S^1_{2,q-\mu},
\end{equation}
$C(j_{1},j_{2},j;m_{1},m_{2},m)$ is the Clebsch-Gordan coefficient,
and $\bm{C}^2$ is the second-rank tensor formed from the
modified spherical harmonics
\begin{equation}
  C^l_m (\theta,\phi) \equiv \sqrt{\frac{4\pi}{2l+1}} \;
  Y_{lm} (\theta, \phi).
\end{equation}  
The radial factor is $V_p(r) = b/r^3$ where
$b \equiv -\sqrt{6}\beta/\hbar^2$.

To second order, the change in energy of the state $|\alpha \rangle $
due to the perturbation $H_{\text{sd}}$ is
\begin{equation}
\label{eq1.9}
\Delta E_\alpha = \Delta E_\alpha^{(1)} + \Delta E_\alpha^{(2)}.
\end{equation}
where
\begin{equation}
\label{eq1.10}
 \Delta E_\alpha^{(1)} = \langle \alpha |H_{\text{sd}}| \alpha \rangle,
\end{equation}
and
\begin{equation}
\label{eq1.11}
\Delta E_\alpha^{(2)} = \sum_{\beta\neq\alpha} 
\frac{\langle \alpha |H_{\text{sd}}|\beta \rangle \langle \beta
H_{\text{sd}}|\alpha \rangle }{E_\alpha^{(0)} - E_\beta^{(0)}}.
\end{equation}
        
The calculation of the first order energy correction is straightforward.
To evaluate the second order correction we use the method of
Dalgarno and Lewis \cite{Dal55} and introduce an operator $\hat{F}$ which
satisfies the inhomogeneous equation
\begin{equation}
\label{eq1.12}
[\hat{F}, H_0] |\alpha \rangle = H_{\text{sd}} |\alpha \rangle,
\end{equation}
so that
\begin{equation}
\label{eq1.13}
\langle \beta |H_{\text{sd}}|\alpha \rangle = ({E_\alpha^{(0)} -
 E_\beta^{(0)}}) \langle \beta |\hat{F}| \alpha \rangle.
\end{equation}
This gives
\begin{equation}
\label{eq1.14}
\Delta E_\alpha^{(2)} = \langle \alpha|H_{\text{sd}} \hat{F} |\alpha \rangle 
 - \langle \alpha| H_{\text{sd}}|\alpha \rangle
\langle \alpha |\hat{F} |\alpha \rangle.
\end{equation}

As the channel states form a complete orthonormal basis over the angular
and electronic coordinates, we can expand the state
$\hat{F}|\alpha \rangle $ in terms of them:
\begin{equation}
\label{eq1.15}
\hat{F}\;|\alpha \rangle = \sum_{a^{\prime}} f_{a^{\prime}}(r)
|\Phi_{a^{\prime}} \rangle,
\end{equation}
so that (\ref{eq1.12}) becomes an inhomogeneous equation for the
radial states $f_{a'}(r)$
\begin{equation}
\label{eq1.16}
\left( E_\alpha^{(0)} - h_{a^{\prime}} \right) f_{a^{\prime}}(r) 
 =  D_{a^{\prime}a}V_p(r)\;r^{-1}\;R_{va}(r)
\end{equation}
where $\langle \Phi_{a^{\prime}}|H_0 |\Phi_{a} \rangle \equiv
h_{a^{\prime}} \delta_{a^{\prime},a}$
and $ \langle \Phi_{a^{\prime}}|H_{\text{sd}}| \Phi_{a}\rangle  \equiv
V_p(r) D_{a^{\prime}a}$.

By invoking the Wigner-Eckart theorem we find that the angular momentum
factors in the matrix elements for the spin-dipole interaction between
the channel states are
\begin{eqnarray}
\label{eq1.17}
D_{a^{\prime}a} & = & \delta_{M_{S^{\prime}} + m^{\prime}, M_S + m}\;
 (-1)^{M_{S^{\prime}} - M_S}  \nonumber  \\*
 && \times  C(S,2,S^{\prime};M_S, M_{S^{\prime}}- M_S,
 M_{S^{\prime}}) \nonumber  \\*
 && \times  C(l,2,l^{\prime}; m, m^{\prime}-m, m^{\prime}) \nonumber \\*
&& \times \langle \Gamma^{\prime} S^{\prime}||\bm{T}^2||\Gamma S \rangle
 \langle l^{\prime}||\bm{C}^2|| l\rangle.
\end{eqnarray}
The reduced matrix elements are given by
\begin{eqnarray}
\label{eq1.18}
 \langle \Gamma^{\prime} S^{\prime}||\bm{T}^2||\Gamma S \rangle & = &
   \delta_{\Gamma^{\prime},\Gamma}\; \delta_{S_1^{\prime},S_1}\;
   \delta_{S_2^{\prime}S_2}\;\hbar^{2} \nonumber  \\*
&& \times  \sqrt{S_1(S_1+1)S_2(S_2+1)}  \nonumber \\*
&& \times   \left[
                        \begin{array}{lll}
                                S_1 & S_2 & S \\
                                1 & 1 & 2 \\
                                S_1 & S_2 & S^{\prime}
                        \end{array}
\right],
\end{eqnarray}
and
\begin{equation}
\label{eq1.19}
\langle l^{\prime}||\bm{C}^2||l \rangle =
\left[ \frac{2l+1}{2l^{\prime}+1} \right]^
 {\frac{1}{2}}\; C(l,2,l^{\prime};0,0,0).
\end{equation}
The angular momentum coefficient in (\ref{eq1.18}) is related to the
Wigner $9-j$ coefficient
\begin{eqnarray}
\label{eq1.20}
\left[  \begin{array}{lll}
                        j_1 & j_2 & j \\
                        k_1 & k_2 & k \\
                        j_1^{\prime} & j_2^{\prime} & j^{\prime}
    \end{array} \right]
 & \equiv &
[(2j_1^{\prime} + 1)(2j_2^{\prime} + 1)(2j + 1)(2k + 1)]^{\frac{1}{2}}
\nonumber  \\*
&& \times   \left\{
                        \begin{array}{lll}
                                j_1 & j_2 & j \\
                                k_1 & k_2 & k \\
                                j_1^{\prime} & j_2^{\prime} & j^{\prime}
                        \end{array}
                \right\}.
\end{eqnarray}

Introducing the functions $g_{a^{\prime}}(r) \equiv (r/b D_{a^{\prime}a})\;
f_{a^{\prime}}(r)$
then (\ref{eq1.16}) becomes the inhomogeneous differential equation
\begin{eqnarray}
\label{eq1.21}
\left[- \frac{\hbar}{2M} \frac{d^2}{dr^2} +
\frac{l^{\prime}(l^{\prime}+1)\hbar^2 }{2M r^2} +
V^{\text{el}}_{\Lambda^{\prime} S^{\prime}}(r) - E_\alpha^{(0)}
\right] {g_{a^{\prime}}}(r) \nonumber  \\*
= -  \frac{1}{r^3} {R_{va}}(r).
\end{eqnarray}

\section{Application to the $v=14$ ${}^{5}\Sigma^{+}_{g}$ state of
H\lowercase{e}(2${}^{3}\text{S}_{1}$)+H\lowercase{e}(2${}^{3}\text{S}_{1}$)}

We now consider application of the theory to the spin-polarized
$s$-wave $v=14$ state in the ${}^5\Sigma_g^+$ potential of two colliding
metastable helium atoms. The
initial unperturbed state is therefore
\begin{equation}
\label{eq2.1}
|\alpha \rangle = r^{-1}R_{v=14,a}(r) |\Phi_a \rangle,
\end{equation}
where 
\begin{equation}
\label{eq2.2}
|\Phi_a \rangle = |\Gamma, S=2, M_S=2 \rangle |l=0, m=0 \rangle.
\end{equation}
Here $\Gamma = \{S_1=1, S_2=1, \bar{\Gamma }\}$ where $\bar{\Gamma}$
denotes the quantum numbers specifying the $\Sigma_g^+$ nature of the
problem.

The spin-dipole interaction coefficients (\ref{eq1.17}) simplify to
\begin{eqnarray}
\label{eq2.3}
 D_{a^{\prime}a} & = & \delta_{M_{S^{\prime}} + m^{\prime}, 2} \;
 \delta_{l^{\prime},2} (-1)^{M_{S^{\prime}}}  \nonumber  \\*
&& \times  C(2,2,S^{\prime};2, M_{S^{\prime}} - 2, M_{S^{\prime}}) \nonumber  \\*
&& \times \langle \Gamma^{\prime} S^{\prime}||\bm{T}^2||\Gamma 2 \rangle
\langle 2||\bm{C}^2||0 \rangle.
\end{eqnarray}
From (\ref{eq2.3}) it is clear that only $d$-waves are connected directly to
the initial $s$-wave state. Consequently the first-order energy shift
for $s$-wave states vanishes. The reduced matrix elements relevant
to the second order calculation are 
$\langle \Gamma^{\prime} 2||\bm{T}^2||\Gamma 2\rangle=
\delta_{\Gamma^{\prime},\Gamma} \;
\hbar^2\sqrt{{7}/{3}}$,
$\langle \Gamma^{\prime} 0||\bm{T}^2||\Gamma 2\rangle= -
\delta_{\Gamma^{\prime},\Gamma}\; \hbar^2 \sqrt{{5}/{3}}$ and
$\langle 2||\bm{C}^2||0\rangle = 1/\sqrt{5}$.  The matrix element
$\langle \Gamma^{\prime} 1||\bm{T}^2||\Gamma 2\rangle$ vanishes by
symmetry of the $9-j$ symbol.

The energy shift for $s$-wave states is given to second order by
\begin{equation}
\label{eq2.4}
 \Delta E_\alpha = \sum_{a^{\prime}} |D_{a^{\prime}a}|^2 \;b^2
 \int_0^{\infty}
 R_{va}^*(r) \frac{1}{r^3} g_{a^{\prime}}(r) dr.
\end{equation}
Since neither $R_{a}(r)$ nor $g_{a^{\prime}}(r)$ depend on the
magnetic quantum
numbers $m$ and $M_S$, we may carry out the summation to obtain finally
\begin{eqnarray}
\label{eq2.5}
 \Delta E_\alpha & = & \frac{2}{5} \beta^2 \int_0^{\infty} R_{va}^*(r)
 \frac{1}{r^3}
   \left[ g_{\Gamma,S^{\prime}=0,l=2}(r) \right. \nonumber  \\*
   && + \left. 7 g_{\Gamma,S^{\prime}=2,l=2}(r) \right] dr.
\end{eqnarray}

\section{Results and Discussion}

The differential equations (\ref{eq1.7}) for the unperturbed $v=14$
${}^{5}\Sigma^{+}_{g}$ bound state and (\ref{eq1.16}) for the
perturbed function $g_{a^{\prime}}(r)$ are solved using a discrete
variable representation and a scaled radial coordinate grid (see
\cite{Peach04} and \cite{Beams05} for details).

As input to the problem we require the Born-Oppenheimer potentials for
both the ${}^5\Sigma_g^+$ and ${}^1\Sigma_g^+$ electronic states of the
metastable helium dimer.  For the ${}^5\Sigma_g^+$ potential we have
used the full (adiabatically and relativistically corrected) potential
$V^{5}_{\text{PJ}}(r)$
of Przybytek and Jeziorski~\cite{Przyb05} adjusted to match the
experimental binding energy of the least bound ($v=14$)
state~\cite{Moal06}.  In order to gauge the sensitivity
to the potentials we have also used the potential $V^{5}_{\text{SM}}(r)$ of
St\"arck and Meyer~\cite{Starck94}, adjusted to match the
experimental value of the least bound state.  For the singlet
potential we use \cite{Venturi99} a potential $V^{1}_{\text{M}}$
constructed from the short-range
M\"uller \textit{et al}~\cite{Muller91} potential  exponentially
damped onto the quintet potential at long range.

In order to model Penning ionization from the singlet state we use a
complex optical potential of the form $V(r)-i\Gamma(r)/2$.  Two forms of
the ionization width $\Gamma(r)$ are used; $\Gamma_{\text{M}}(r)$ obtained
from a least squares fit to the
tabulated results in~\cite{Muller91} and the simpler form
$\Gamma_{\text{GMS}}(r)=0.3 \exp(-r/1.086)$ advocated
in~\cite{Garrison73}.  The sensitivity of the calculation to the
ionization widths was estimated by assigning a 20\% relative uncertainty
to both forms of the width.

The coupling between the initial quintet state and the singlet state
produces a complex energy shift
$\Delta E_\alpha = \Delta E_\alpha^{\text{re}} - i\gamma_\alpha/2$
due to the complex form of the singlet potential.
The $1/e$ lifetime of
the least bound $v=14$ state of the ${}^5\Sigma_g^+$ potential is then
$\tau =\hbar/\gamma_\alpha$.  Using the adjusted
Przybytek and Jeziorski quintet potential~\cite{Przyb05} we obtain
(see Table \ref{lifetimes})
values of $\tau = (1.578 ^{+0.201}_{-0.116}) \times 10^{-4}$ s and
$\tau = (1.376 ^{+0.108}_{-0.055}) \times 10^{-4}$ s,
using the two forms of the
ionization width~\cite{Muller91} and~\cite{Garrison73} respectively.
The superscripted (subscripted) numbers indicate the uncertainty due to
a negative (positive) variation of 20\% in the widths.
The use of the modified St\"arck and Meyer quintet
potential changes the lifetime (see Table \ref{lifetimes}) by
$\alt 2\,\mu$s.
We have also studied the sensitivity of the lifetime to realistic variations
in the singlet potential. We find that (see Table \ref{lifetimes}) a
variation of $+2\% $ changes the lifetime by $\alt 9\,\mu$s (a variation
of $-2\% $ was rejected as it reduces the number of $d$-wave bound
states from 28 to 27).
We note that, for all cases, the real part of the energy shift
increases the binding energy by about 10 kHz.

\begin{table}
\caption{\label{lifetimes} Spin-dipole induced lifetime
(in units of $10^{-4}$ s) of the $v=14$
${}^{5}\Sigma^{+}_{g}$ state. Results are shown for the two
quintet molecular potentials
$V^{5}_{\text{PJ}}(r)$ and $V^{5}_{\text{SM}}(r)$, and
the two ionization widths
$\Gamma_{\text{M}}(r)$ and $\Gamma_{\text{GMS}}(r)$. Also
shown are results for $\pm 20\%$ variation in the ionization widths. 
Results in brackets correspond to a $+2\%$ variation in the singlet
potential $V^{1}_{\text{M}}(r)$.}
\begin{ruledtabular}
\begin{tabular}{lccc}
& $\Gamma_{\text{M}}$ & $0.8\Gamma_{\text{M}}$
& $1.2\Gamma_{\text{M}}$ \\
\hline
 $V^{5}_{\text{PJ}}$ & 1.578 (1.643) & 1.779 (1.870) & 1.462 (1.510)    \\
 $V^{5}_{\text{SM}}$ & 1.597 (1.630) & 1.806 (1.850) & 1.475 (1.501)     \\
&&&\\
& $\Gamma_{\text{GMS}}$
& $0.8\Gamma_{\text{GMS}}$
& $1.2\Gamma_{\text{GMS}}$  \\
\hline
 $V^{5}_{\text{PJ}}$ & 1.376 (1.429) & 1.484 (1.556) & 1.321 (1.359)   \\
 $V^{5}_{\text{SM}}$ & 1.386 (1.424) & 1.499 (1.546) & 1.327 (1.357)    \\

\end{tabular}
\end{ruledtabular}
\end{table}

Due to the disparity between our calculated value of around
150 $\mu$s and the observed experimental lifetime of a
few $\mu$s~\cite{Moal06}, we considered if there were
other spin-dipole induced processes which may contribute to the lifetime.
It is our belief that the only possible energy-conserving transitions
from the initial bound state are to the ionization continuum via the singlet
state.  For example the $d$-wave scattering states are much closer
than the nearest lying $d$-wave bound state in either the singlet or
quintet potential, but have negligible overlap with the $v=14$ state
due to the centrifugal barrier for $l=2$.  Due to the selection rule
$M_{S^{\prime}} + m^{\prime} = M_S + m = 2$, second-order transitions
back to $s$-wave states with $M_S<2$ are forbidden.  We believe that
the observed lifetime is not spin-dipole induced but due to some other
process such as exchange of vibrational and translational energy
through atom-molecule collisions.

\end{document}